\documentclass[letterpaper, 10 pt, conference]{ieeeconf}  

\IEEEoverridecommandlockouts                              

\usepackage{graphics, comment} 
\usepackage{epsfig} 
\usepackage{times} 
\usepackage{amsmath} 
\usepackage{amssymb}  
\usepackage{algorithm}
\usepackage{xcolor}
\usepackage{amsfonts}
\usepackage{subcaption}
\usepackage{graphicx}
\usepackage{algorithmic}
\usepackage{mathtools} 
\usepackage{hyperref}
\usepackage{bm}  
\usepackage{cleveref}
\usepackage{eso-pic}
\usepackage{graphicx}
\usepackage{tikz}
\usepackage{pgfplots}
\pgfplotsset{compat=1.18}
\usetikzlibrary{positioning, shapes.geometric}
\usetikzlibrary{arrows.meta}

\AddToShipoutPictureFG*{%
  \put(\LenToUnit{0.5\paperwidth},\LenToUnit{0.55in}){%
    \makebox(0,0){\parbox{0.85\paperwidth}{\centering\scriptsize
      Accepted to the 2026 IEEE Conference on Decision and Control (CDC).\\
      \copyright~2026 IEEE. Personal use of this material is permitted.
      Permission from IEEE must be obtained for all other uses, in any
      current or future media, including reprinting/republishing this
      material for advertising or promotional purposes, creating new
      collective works, for resale or redistribution to servers or lists,
      or reuse of any copyrighted component of this work in other works.}}}%
}

\crefname{section}{Section}{Sections}
\crefname{algorithm}{Algorithm}{Algorithms}
\crefname{figure}{Fig.}{Fig.}
\newtheorem{proposition}{Proposition}
\newtheorem{remark}{Remark}
\crefname{proposition}{Proposition}{Proposition}

\title{\LARGE \bf
Dynamic Multi-Robot Task Allocation under Uncertainty and Communication Constraints: A Game-Theoretic Approach
}

\author{Maria G. Mendoza$^{*1}$, Pan-Yang Su$^{*2}$, Bryce L. Ferguson$^{3}$, S. Shankar Sastry$^{2}$
\thanks{This work was supported in part by the following: Provably Correct Design of Adaptive Hybrid Neuro-Symbolic Cyber Physical Systems, DAF Air Force Research Laboratory award number FA8750-23-C-0080; Stress Testing and Hardening the NAS for Safe, Efficient, and Resilient Growth, NASA National Aeronautics and Space Administration award number 80NSSC24M0068; Collaborative Research: Transferable, Hierarchical, Expressive, Optimal, Robust, Interpretable NETworks (THEORINET) under Simons Foundation Award No. MPS-MODL-00814647 and NSF Award No. 2031899.}
\thanks{*Equal contribution and listed alphabetically}
\thanks{$^{1}$Department of Mechanical Engineering, University of California, Berkeley, Berkeley, CA, United States
        {\tt\small maria\_mendoza@berkeley.edu}}%
\thanks{$^{2}$Department of Electrical Engineering and Computer Science, University of California, Berkeley, Berkeley, CA, United States
        {\tt\small pan\_yang\_su@berkeley.edu, sastry@coe.berkeley.edu}}%
\thanks{$^{3}$Thayer School of Engineering, Dartmouth College, Hanover, NH, United States
        {\tt\small bryce.l.ferguson@dartmouth.edu}.}%
}

\begin{document}

\maketitle
\thispagestyle{empty}
\pagestyle{empty}

\begin{abstract}
We study dynamic multi-robot task allocation under uncertain task completion, time-window constraints, and incomplete information. Tasks arrive online over a finite horizon and must be completed within specified deadlines, while agents operate from distributed hubs with limited sensing and communication. We model incomplete information through hub-based sensing regions that determine task visibility and a communication graph that governs inter-hub information exchange. Using this framework, we propose Iterative Best Response (IBR), a decentralized policy in which each agent selects the task that maximizes its marginal contribution to the locally observed welfare. We compare IBR against three baselines: Earliest Due Date first (EDD), Hungarian algorithm, and Stochastic Conflict-Based Allocation (SCoBA), on a city-scale package-delivery domain with up to 100 drones and varying task arrival scenarios. Under full and sparse communication, IBR achieves competitive task-completion performance with lower computation time.

\end{abstract}

\section{Introduction}
Multi-robot task allocation (MRTA) is used in many applications, including package delivery \cite{dynamic-mr-ta}, disaster response \cite{GHASSEMI2022103905}, warehouse dispatch \cite{8460886}, and more. In many of these settings, tasks arrive sequentially over a finite planning horizon, must be completed within specified time windows, and face uncertain completion outcomes due to stochastic travel times and environmental variability~\cite{dynamic-mr-ta}. Moreover, agents often operate under \emph{incomplete information}: limited sensing ranges restrict task visibility, and communication constraints prevent agents from observing others' decisions~\cite {Grimsman2-valid}. 

While MRTA problems have been studied extensively across diverse problem classes, coordination architectures, and solution methods, existing approaches are insufficient to address the combined challenges of online task arrivals, spatio-temporal constraints, and uncertainty in task completion while operating with incomplete information and communication constraints. In this work, we consider these challenges and design a decentralized policy that maximizes the total number of tasks completed within their respective windows. Following the taxonomy of~\cite{CHAKRAA2023104492}, our problem belongs to the class of ST-SR-TA with online assignment: each agent executes one task at a time (single-task, ST), each task requires one agent (single-robot, SR), and agents plan task sequences over a finite horizon subject to time-window constraints (time-extended assignment, TA).
 
Decentralized MRTA methods typically assume static task environments and do not scale well as the number of agents or tasks increases \cite{CHAKRAA2023104492}. Conversely, dynamic MRTA approaches \cite{dynamic-mr-ta,GHASSEMI2022103905,liu2026learning} frequently rely on centralized computation or global coordination signals, thus limiting their applicability to large-scale, distributed robotic teams. Our work bridges this gap by addressing dynamic MRTA under decentralized execution with explicit communication constraints.

\subsection{Proposed Model for Incomplete Information}
\label{subsec: proposed model}
We introduce three modeling constructs to capture incomplete information as it arises in practice, drawing on ideas from distributed optimization and valid utility games~\cite{Grimsman2-valid,grimsman-2019-impact}. \emph{Hubs} partition agents according to operational structure (e.g., depot locations or sector assignments). \emph{Task visibility} restricts which tasks each agent can observe based on spatial sensing regions tied to its assigned hub. The \emph{communication graph} over hubs specifies which hubs exchange task decision information during planning and execution. Designing this grouping structure introduces a fundamental trade-off: richer information sharing can improve system-level task performance, but it incurs increased communication and computational costs. Understanding how this structure impacts system-level performance is a key objective of this work.



\subsection{Proposed Algorithm for Decentralized Task Allocation}
\label{subsec: IBR}
Game-theoretic coordination mechanisms remain rare in decentralized MRTA problems, appearing in fewer than 3\% of reviewed works \cite{CHAKRAA2023104492}.
Under our model, we propose Iterative Best Response (IBR) as a simple, decentralized policy in which each agent selects the task that maximizes its marginal contribution to the locally observed welfare. We evaluate how efficiency degrades as communication becomes sparse, delayed, or highly restrictive, and we demonstrate scalable coordination without centralized control.

Furthermore, we compare IBR against three baselines: Earliest Due Date first (EDD) \cite[Chapter 3.2]{10.5555/1477600}, Hungarian algorithm \cite{doi:10.1137/0105003}, and Stochastic Conflict-Based Allocation (SCoBA) \cite{dynamic-mr-ta}, using simulations for multi-drone package delivery. Our results show that IBR achieves superior performance while maintaining high computational efficiency. 

\textit{\textbf{Our contributions are:}} (1) A modeling framework for decentralized dynamic MRTA that captures incomplete information through hub-based sensing regions and inter-hub communication graphs; (2) IBR as a decentralized allocation policy where each agent picks the task maximizing its marginal contribution to locally observed welfare; (3) Extensive empirical evaluation across a range of operational conditions: task load, service-window duration, fleet configuration, and spatial conflict density, as well as communication graph topologies, showing that IBR consistently matches or outperforms centralized and decentralized baselines in task-completion rate and computation time; (4) Efficiency analysis under varying communication graph structures and operational conditions.

\section{Related Works}

\textbf{Centralized dynamic MRTA.}
Most approaches to dynamic MRTA rely on centralized or hierarchical deconfliction mechanisms, or assume that task assignments are globally broadcast immediately after decisions. Consequently, these models often fail to capture the combination of execution uncertainty, limited sensing, and strict communication constraints that necessitate fully decentralized execution. Lerman et al.~\cite{doi:10.1177/0278364906063426} provided an early analytical model but assumed centralized information. More recent work has integrated complex operational constraints: Choudhury et al.~\cite{dynamic-mr-ta} address execution uncertainty and time windows, and Ghassemi and Chowdhury~\cite{GHASSEMI2022103905} incorporate task deadlines with robot range and payload limits. Tuck et al.~\cite{tuck2024smt} handle online task arrivals with deadlines, using Satisfiability Modulo Theories (SMT) to guarantee soundness and completeness.  However, these all depend on centralized conflict resolution or global task visibility.  Similarly, while Liu et al. \cite{liu2026learning} recently optimized spatio-temporal allocations using learned trait-efficacy maps, their framework assumes deterministic execution, perfect communication, and a fixed task set. In contrast, our work jointly handles dynamic task arrivals, probabilistic completion, and sparse communication under decentralized execution.

\textbf{Distributed MRTA and communication-aware methods.}
Various distributed MRTA approaches have been proposed to move beyond centralized setups, ranging from classic Hungarian methods \cite{giordani2010distributed} to modern global games \cite{kanakia2016modeling,beaver2025globalgamesinspiredapproachmultirobot}, hedonic coalition formation \cite{czarnecki2021scalable}, market-based mechanisms \cite{9341652,10847745}, and reinforcement learning \cite{9981353}. While effective in specific contexts, these models often rely on simplifying assumptions that neglect deadline constraints and the heterogeneous sensing capabilities of individual agents. Furthermore, many of these approaches still require global coordination signals to synchronize agent actions. Several works integrate communication graphs to formally model information exchange \cite{5072249, giordani2010distributed, 10847745, 9360591}. However, a distinction exists between our approach and existing work: whereas prior literature often focuses on ad hoc robot-to-robot communication, our model operates at the multi-depot level. More importantly, these existing graph-based methods still fail to jointly account for the probabilistic execution and temporal constraints.
 
\textbf{Distributed methods for dynamic MRTA.}
Recent work has begun to consider distributed approaches within dynamic MRTA, but remains limited in scope. Chen et al. \cite{CHEN201931} propose a distributed method with critical time constraints, yet their notion of dynamics concerns changes in the communication graph rather than online task arrivals. Similarly, Ghassemi et al. \cite{8901062} address decentralized MRTA with task deadlines, range limits, and asynchronous decision-making under dynamic task spaces, but assume deterministic success probabilities and require a subset of global information to be available to all robots. Our framework removes both of these limitations by operating under stochastic task completion with strictly local information.

\section{Problem Formulation} 
\label{problem-form}

We formulate our model in \cref{subsec: model} and discuss motivating applications in \cref{subsec: applications}.

\subsection{Model}
\label{subsec: model}
We consider a finite set of agents $\mathcal{N} = \{1,2,\dots,n\}$ operating on a metric space $(X, d)$ over a discrete time horizon $\mathcal{T} = \{1,2,\dots,T\}$.\footnote{In this work, $(X, d)$ is usually taken as $\mathbb{R}^2$ with the Euclidean norm.} Agents are organized around a finite set of hubs (or depots) $\mathcal{H} = \{1,2,\dots,H\}$. Each hub $h \in \mathcal{H}$ is associated with a fixed spatial location $L_h \in X$ and a sensing region $R_h \subset X$, and every agent $i \in \mathcal{N}$ belongs to exactly one hub, denoted by $\eta(i) \in \mathcal{H}$. One example of a sensing region is a ball centered at the hub's location; that is, $R_h =\{x \in X: d(L_h, x) \le r\}$ for some radius $r > 0$. 

The hub assignment $\eta(\cdot)$ is fixed over the planning horizon and represents an agent’s home base for sensing, coordination, and deployment. While agents may move throughout the environment during task execution, their sensing and task awareness are determined by their assigned hub.

\paragraph{\textbf{Tasks}} A finite set of tasks $\mathcal{K}$ arrives over time. Each task $k \in \mathcal{K}$ is characterized by a tuple $(a_k, \underline{t}_k, \overline{t}_k, \ell_k)$, where $a_k \in \mathcal{T}$ denotes the reveal (arrival) time, $[\underline{t}_k,\overline{t}_k]\subset \mathcal{T}$ with $a_k \le \underline{t}_k \le \overline{t}_k$ denotes the admissible \emph{service} window, and $\ell_k \in X$ denotes the location. A task $k$ is said to be \emph{visible} to hub $h$ if $\ell_k \in R_h$. We define the set of tasks that agent $i$ can observe at time $t$  as $\mathcal{K}_{it} = \{k \in \mathcal{K}: \ell_k \in R_{\eta(i)}, a_k \leq t \}$.

\begin{figure}
    \centering
    \includegraphics[width=1\linewidth]{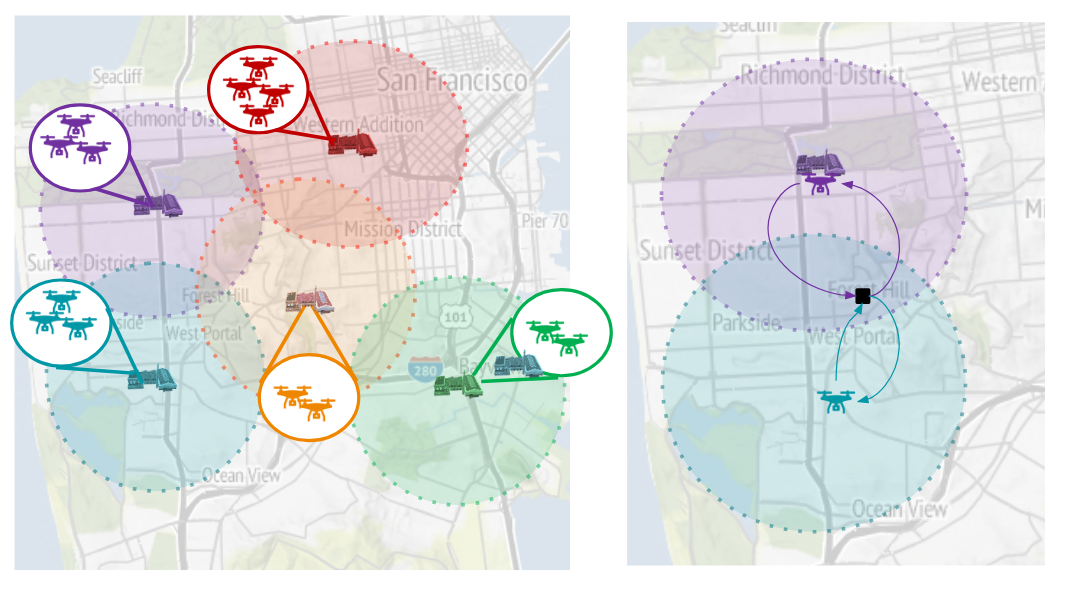}
    \caption{Decentralized multi-hub architecture. Each hub acts as a local planner coordinating its assigned drones. Shaded regions represent hub-based sensing areas that determine task visibility. Overlapping sensing regions imply that certain tasks are visible to multiple hubs.
}
    \label{fig:Map-setup}
\end{figure}

Task locations may lie within multiple sensing regions (see \cref{fig:Map-setup}), so the same task can be visible to agents at different hubs. This overlap creates potential competition and redundancy in allocations when communication is limited.

To capture stochastic task completion, we adopt the uncertainty model of \cite{dynamic-mr-ta}, in which uncertainty in task execution arises from stochastic travel times. For each hub $h$ and task $k$, we let $\tau_{hk} \sim F_{hk}$ denote the stochastic travel time\footnote{The forward and return travel times are sampled from the same distribution, but their realizations may be different.} between hub $h$ and task $k$ with cumulative distribution function $F_{hk}$ parametrized by the distance $d(L_h, \ell_k)$. If agent $i$ (with hub $h=\eta(i)$) departs its hub at time $t$ to service task $k$, then its arrival time is $t+\tau_{hk}$, and the task is said to be \emph{completed} if the agent arrives within the service window, i.e., $\underline{t}_k \le t+\tau_{hk} \le \overline{t}_k$. Note that if the agent arrives at the task early ($t+\tau_{hk} < \underline{t}_k$), we assume it waits until the service window begins, so the task is still completed. Thus, the success probability $p_{ik}(t)$ is given as
\begin{equation} \label{eq:success-prob}
p_{ik}(t) = \mathbb{P} [i \ \text{completes} \ k \ | \ t] = F_{\eta(i)k}(\overline{t}_k - t)
\end{equation}
We let $v_k(t) \in\{0, 1\}$ be the indicator variable such that $v_k(t) = 1$ if and only if task $k$ is completed no later than time $t$. Also, we require that agents return to their assigned hub after servicing a task before attempting a new task, and the return time is given by $t^{\mathrm{return}} = t + \tau_{hk}^f + \tau_{hk}^b$. We define $\tau_{hk}^f = \tau_{hk} + \max(\underline{t}_k - t - \tau_{hk}, 0)$ to account for the possibility of arriving earlier and waiting, and $\tau_{hk}^b$ is independent of $\tau_{hk}$ with distribution $\tau_{hk}^b \sim F_{hk}$. 

\paragraph{\textbf{Agents}}
At each time step $t \in \mathcal{T}$, an agent $i \in \mathcal{N}$ is characterized by the tuple $(p_i(t), a_i(t), I_i(t))$, where $p_i(t) \in X$ represents its location, $a_i(t) \in A_i(t)$ its action, and $I_i(t)$ its local information set. We describe each component below.

First, $p_i(t)$ denotes the agent's current physical location. Specifically, $p_i(t) = L_{\eta(i)}$ indicates that the agent is at its assigned hub. Otherwise, the agent is actively executing a task or returning to the hub.
Second, the action $a_i(t)$ is chosen from the feasible set $A_i(t) = \mathcal{K}_{it} \cup \{\rho, \varnothing\}$. This set includes three operational modes: an \emph{execution mode}, in which the agent selects a visible task $k \in \mathcal{K}_{it}$ to pursue, a \emph{return mode}, in which the agent is returning to the hub (denoted by $\rho$), and an \emph{idle mode} (denoted by $\varnothing$), in which the agent remains at the hub available for new assignments.

A new task may be selected only while idle, i.e., $a_i(t) \in \mathcal{K}_{it} \Rightarrow a_i(t-1) = \varnothing$ (with $a_i(0) = \varnothing$ for all $i$). In addition, we impose the following constraints on the evolution of the locations and actions of the agents.
\begin{enumerate}
    \item The location evolves according to a discrete-time dynamic $f$, which will be explicitly defined in \cref{sec:simulation}.
    \begin{equation}
    \label{eq: constraint 1}
    p_i(t+1) = f(p_i(t), a_i(t))
    \end{equation}
    
    \item Once an agent leaves the hub, it remains committed to the task (execution mode). Once the agent arrives at the task location, or the task has expired, it returns to the hub (return mode).
    \begin{equation}
    \label{eq: constraint 2}
    \begin{aligned}
    &(a_i(t) = k) \land\! \neg\left[(p_i(t+1) = \ell_k) \lor (t+1 > \bar{t}_k)\right] \\
    &\Rightarrow a_i(t+1) = k
    \end{aligned}
    \end{equation}
    \begin{equation}
    \label{eq: constraint return}
    \begin{aligned}
    &(a_i(t) = k) \land \left[(p_i(t+1) = \ell_k) \lor (t+1 > \bar{t}_k)\right] \\
    &\Rightarrow a_i(t+1) = \rho
    \end{aligned}
    \end{equation}
    
    \item A returning agent remains in the return mode until it reaches the hub. After that, it switches to the idle mode.
    \begin{equation}
    (a_i(t) =\rho) \land (p_i(t+1) \neq L_{\eta(i)}) \Rightarrow a_i(t+1) = \rho
    \end{equation}
    \begin{equation}
    \label{eq: constraint 3}
    (a_i(t) =\rho) \land (p_i(t+1) = L_{\eta(i)}) \Rightarrow a_i(t+1) = \varnothing
    \end{equation}
\end{enumerate}

Information sharing is governed by a directed communication graph $\mathcal{G} = (\mathcal{H}, \mathcal{E})$, where $(h_1,h_2) \in \mathcal{E}$ indicates that agents in hub $h_1$ can observe the actions or intentions of agents in hub $h_2$.  The graph may be sparse or asymmetric.  

Agents’ decisions are thus based solely on locally visible tasks and the observed actions of agents within their communication neighborhood (including agents from the same hub) $\mathcal{N}_i = \{j \in \mathcal{N}: (\eta(i), \eta(j)) \in \mathcal{E} \text{ or } \eta(i) = \eta(j)\}$. The information available to agent $i$ at time $t$ is
$$
\begin{aligned}
I_i(t) \!= \mathcal{K}_{it} \cup \{a_j(s) \!: j \in \mathcal{N}_i, s < t\} \cup \{x_j(t) \!: j \in \mathcal{N}_i\},
\end{aligned}
$$
where $x_j(t)$ is the temporary chosen action by agent $j$ at time $t$ as formally described in \cref{sec:ibr}.  

With $I_i(t)$, we let $\mathcal{K}_{it}'$ represent the subset of tasks observable to agent $i$ that have not been attempted before, based entirely on its local information.
\begin{equation}
\label{eq: available}
\begin{aligned}
\mathcal{K}_{it}' = &\{k \in \mathcal{K}_{it}: \bar{t}_k \geq t\} \setminus \\
&\{a_j(s): j \in \mathcal{N}_i, s < t, a_j(s) \neq \varnothing\}
\end{aligned}
\end{equation}

Task allocations made at earlier times are not globally observable. An agent does not, in general, retain knowledge of which tasks have been previously assigned to other agents unless such assignments are revealed through its communication neighborhood. In particular, if an agent $i$ and an agent $j$ do not share a communication edge in $\mathcal{G}$, then agent $i$ cannot observe past or current task selections made by agent $j$, even if those selections were made at earlier time steps or are currently being attempted. An agent learns whether a task has already been completed only upon arriving at the task location.

Before execution, task allocations remain local decisions that may be revised as new tasks arrive, time windows evolve, or additional information becomes available. Our work differs from existing dynamic MRTA work \cite{doi:10.1177/0278364906063426,dynamic-mr-ta,GHASSEMI2022103905,liu2026learning} in this modeling approach, where the assumption is that assignments become globally known after assignment.

\paragraph{\textbf{Optimization Problem and Policies}}
We seek a policy that maximizes the expected number of completed tasks. A policy $\pi: (i, t, p_i(t), I_i(t)) \mapsto a_i(t)$ describes agent $i$'s action at time step $t$ based only on the current location $p_i(t)$ and information $I_i(t)$ that is locally available at time $t$, such that the constraints \eqref{eq: constraint 1}-\eqref{eq: constraint 3} are satisfied. Thus, our optimization problem is 
\begin{equation}
\label{eq: opt}
\begin{aligned}
&\max_{\pi} \quad \mathbb{E}[\sum_{k \in \mathcal{K}} v_{k}(T) | \pi]\\
&\ \text{s.t.} \quad \eqref{eq: constraint 1}-\eqref{eq: constraint 3}
\end{aligned}
\end{equation}

Note that under this policy, agents commit to one task at a time, rather than a sequence of tasks.

\subsection{Applications}
\label{subsec: applications}
Our model is motivated by several real-world applications. For example, companies operating drone-based package delivery services must coordinate fleets of autonomous vehicles under time windows, uncertain travel times, and large-scale deployments. While execution-level coordination (e.g., collision avoidance) may rely on shared airspace information, task allocation decisions may be distributed across regional hubs or operational sectors for scalability and resilience. In dense urban environments with thousands of concurrent deliveries, fully centralized coordination can become computationally and communication expensive. Moreover, when multiple service providers operate in shared airspace, task-level information is not shared globally across companies. As a result, planning decisions are often made under partial and localized visibility, which can lead to redundant or conflicting allocations. 

In disaster response, autonomous vehicles or drones may be tasked with delivering medical supplies, food and water, assessing infrastructure damage, or locating survivors under severe time pressure. Communication infrastructure is often degraded, intermittent, or geographically fragmented, forcing agents to coordinate under localized and incomplete information. In such settings, efficient allocation and distribution of limited resources must be achieved despite uncertainty in travel times and strict service deadlines. Designing scalable coordination mechanisms under sparse communication is therefore essential for resilient emergency response systems.

\section{Iterative Best Response (IBR)} 
\label{sec:ibr}

For each time step $t$, our optimization problem \eqref{eq: opt} can be formulated as an instance of the vehicle-target assignment problem in \cite{grimsman-2019-impact}. Let $x = (x_j)_{j\in \mathcal{N}}$ denote the action for all the agents, and let $\mathcal{K}_t := \bigcup_{i \in \mathcal{N}} \mathcal{K}_{it}$ denote the set of tasks visible to at least one agent at time $t$. The social welfare is thus given as
\begin{equation}
\label{eq: W(x, t)}
W(x,t) =
\sum_{k \in \mathcal{K}_{t}} \left(1 - \prod_{j \in \mathcal{N}: x_j = k} (1 - p_{jk}(t)) \right)
\end{equation}
This formulation means that a task is considered completed as soon as at least one assigned agent completes it. The maximization of $W(x, t)$ is a submodular optimization problem, so a greedy algorithm yields good performance \cite{grimsman-2019-impact}.

\subsection{Iterative Best Response with Local Information}
To formalize the algorithm, we introduce some notation. Under the action profile $x$, for each agent $i$, we let $W_{i}(x,t)$ denote the local welfare it can observe.
\begin{equation}
\label{eq: Wi(x, t)}
W_{i}(x,t) =
\sum_{k \in \mathcal{K}'_{it}} \left(1 - \prod_{j \in \mathcal{N}_i: x_j = k} (1 - p_{jk}(t)) \right)
\end{equation}

Let $(k, x_{-i})$ denote the action profile in which agent $i$ selects task $k$, and $(\varnothing, x_{-i})$ denote the profile where agent $i$ remains idle. Agent $i$ evaluates the marginal contribution of selecting task $k$ as the difference in observed local welfare:
\begin{equation}
\label{eq: marginal}
U_i((k, x_{-i}), t) = W_i((k, x_{-i}), t) - W_i((\varnothing, x_{-i}), t)
\end{equation}
We can interpret each agent's local welfare contribution as its utility, so that the allocation procedure can be viewed as a sequential best-response process in a valid utility game with information constraints \cite{Grimsman2-valid}. At any time $t$, each agent $i \in \mathcal{N}$ locally chooses an action $a_i(t)$ that maximizes its utility. Given its local information $I_i(t)$, agent $i$ first determines the set of available tasks according to \eqref{eq: available}. The IBR update rule for agent $i$ is therefore given by:
\begin{equation}
\label{eq: max ait}
x_i \in \arg\max_{k \in \mathcal{K}_{it}' \cup \{\varnothing\}}
U_i((k, x_{-i}), t)
\end{equation}

Intuitively, this decision rule dictates that each agent selects the task yielding the greatest marginal increase to the collective welfare of its observable neighborhood.

\subsection{Algorithmic Procedure}
Let $\mathcal{N}^{\text{idle}} \leftarrow \{i \in \mathcal{N} : a_i(t-1) = \varnothing\}$ be the set of idle agents that can be assigned new tasks. Our IBR policy has three stages.
\begin{enumerate}
    \item \textbf{Initialization:} Each idle agent $i \in \mathcal{N}^{\text{idle}}$ determines the locally observable set of uncompleted tasks $\mathcal{K}_{it}'$ and computes the corresponding success probabilities $p_{ik}(t)$ for all $k \in \mathcal{K}_{it}'$.
    \item \textbf{Iterative Optimization:} Agents sequentially update their task selections to maximize their marginal utility by solving \eqref{eq: max ait}. We denote the realized order by $\sigma$. This best-response loop continues until no agent can strictly improve its utility or the maximum number of rounds $\bar{r}$ is reached.
    \item \textbf{Action Commitment:} Agents in $\mathcal{N}^{\text{idle}}$ set their actions according to the converged profile $x$. For all other agents, their actions follow from \eqref{eq: constraint 2}-\eqref{eq: constraint 3}.
\end{enumerate}

We provide more details for the second stage. The agents can update at any ordering $\sigma$, so there is no need for synchronization. However, we assume that no agents update at the same time to alleviate ambiguity in computing \eqref{eq: Wi(x, t)}. 

The computational complexity of our algorithm is $O(|\mathcal{K}|n\bar{r})$, which is linear in the number of agents and the number of tasks. In our simulation in \cref{sec:simulation}, setting $\bar{r} = 2$ often suffices.

\subsection{Theoretical Analysis}
We provide a performance guarantee for our algorithm in the single-round case ($\bar{r} = 1$), following the results in \cite{grimsman-2019-impact}. Let $x^*$ be an optimal assignment that solves \eqref{eq: W(x, t)} over $\mathcal{K}_t$. We aim to lower bound $\frac{W(x, t)}{W(x^*, t)}$ based on the structure of the communication graph $\mathcal{G}$.

In \cite{grimsman-2019-impact}, the authors analyze a sequential decision process in which agents act in some order, and each agent's information is restricted to agents that decided \emph{before} it. The direction of information flow of our IBR is dictated by the communication graph since all agents belonging to the same hub can communicate with each other.

Given a communication graph $\mathcal{G} = (\mathcal{H}, \mathcal{E})$, a \emph{clique} is a set of vertices $C \subseteq \mathcal{H}$ such that for every $i, j \in C$, both $(i, j) \in \mathcal{E}$ and $(j, i) \in \mathcal{E}$.\footnote{Our definition is stricter than that in \cite{grimsman-2019-impact}, which only requires that either $(i, j) \in \mathcal{E}$ or $(j, i) \in \mathcal{E}$. The reason stems from our more general formulation. We do not fix an ordering $\sigma$ in the second stage of IBR. In contrast, in \cite{grimsman-2019-impact}, there is an imposed ordering $\prec$ on the vertices such that $(i, j) \in \mathcal{E}$ implies $i \prec j$.} Let $K(\mathcal{G})$ be the set of all cliques in $\mathcal{G}$, and let $M \in \mathbb{R}^{|K(\mathcal{G})|
\times H}$ be the binary matrix whose rows are indicator vectors for the cliques in $\mathcal{G}$; that is, $M_{ij} = 1$ if vertex $j$ belongs to clique $i$ in $\mathcal{G}$, and $0$ otherwise. The fractional independence number $\alpha^*(\mathcal{G})$ is then defined as
\begin{equation}
\label{eq: indep}
\begin{aligned}
\alpha ^*(\mathcal{G}) := \max_{z \in \mathbb{R}^H} & \qquad z^T {\bf 1}\\
\text{s.t.} & \qquad Mz \leq {\bf 1} \\
& \qquad z \geq {\bf 0}.
\end{aligned}
\end{equation}
Intuitively, an independent set is a group of hubs no two of which share an edge, so they cannot communicate. The independence number is the size of the largest such group and can be characterized by \eqref{eq: indep} when $z$ is constrained to be integer-valued. The fractional independence number relaxes this integer constraint. By \cite[Theorem 1]{grimsman-2019-impact}, the performance ratio
$\frac{W(x, t)}{W(x^*, t)}$ is lower bounded by
$\frac{1}{\alpha^*(\mathcal{G})+ 1}$. Specifically, under full communication, $\alpha^*(\mathcal{G}) = 1$, recovering the well-known competitive ratio of $1/2$ for greedy submodular maximization.

\begin{proposition}
\label{prop: 1}
For any communication graph $\mathcal{G}$ and any realized update order $\sigma$, if $\bar{r} = 1$, then 
$$
W(x, t) \geq \frac{1}{\alpha^*(\mathcal{G})+ 1}W(x^*, t).
$$
\end{proposition}

\begin{remark}
Since agents within a hub always communicate, $\alpha^*(\mathcal{G})$ computed on the hub-based communication graph coincides with the
fractional independence number of the induced agent-level graph $\mathcal{G}_A = (\mathcal{N}, \mathcal{E}_A)$ with $(i,j) \in \mathcal{E}_A \iff (\eta(i), \eta(j)) \in \mathcal{E}$ (proof omitted for space); we therefore compute $\alpha^*(\mathcal{G})$ directly on $\mathcal{G}$.
\end{remark}

The above guarantee is for a single time step and $\bar{r} = 1$. The whole optimization problem \eqref{eq: opt} is more difficult to analyze due to the consideration of future task arrivals. As we will show in \cref{sec:comms-results}, our IBR algorithm still achieves good performance under this general setting.




\section{Simulation Methodology}
\label{sec:simulation}

We evaluate the proposed framework in the multi-drone package-delivery domain adapted from~\cite{dynamic-mr-ta}.
The simulation models a fleet of autonomous drones dispatched from multiple depots to deliver packages across a city-scale environment under time-window constraints, uncertain travel times, and the communication and task-visibility constraints introduced in \cref{problem-form}.
Our implementation extends~\cite{dynamic-mr-ta} by incorporating the communication graph framework and the IBR policy described in \cref{problem-form,sec:ibr}.\footnote{We extend the codebase of~\cite{dynamic-mr-ta} from Julia to Python.}
A key distinction from~\cite{dynamic-mr-ta} is that conflict resolution is no longer centralized: once a drone selects a package, drones at depots outside its communication group are unaware of the assignment. Consequently, if two agents from different communication groups independently select the same task, neither realizes the duplication until one physically arrives at the delivery location and claims it.

The environment is based on the North San Francisco delivery scenario from~\cite{dynamic-mr-ta}, spanning approximately 150~km$^2$. Depots are placed at fixed geographic coordinates across the city, and drones are distributed equally across depots. Each drone has a maximum service range of $r = 5$~km from its assigned depot, so the sensing region introduced in \cref{problem-form} takes the form $R_h = \{x \in X : d(L_h, x) \le r\}$.

\subsection{Simulation Setting}
\paragraph{\textbf{Task generation}}
 \begin{figure}
 \vspace{0.2cm}
     \centering
     \includegraphics[width=1\linewidth]{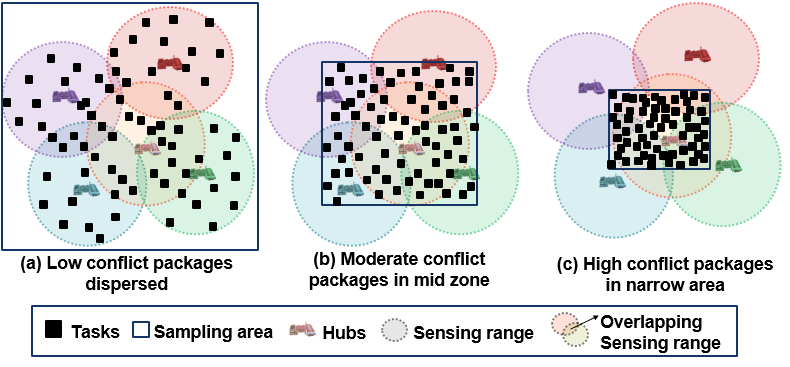}
     \caption{Package generation zones of varying size controlling task conflict density. (a) A large generation zone disperses tasks across the environment, reducing hub overlap and inter-agent conflict, but increasing expected travel times and task completion uncertainty. (b) A medium zone concentrates tasks in a region visible to multiple hubs, producing a moderate level of conflict and intermediate travel distances. (c) A small zone places nearly all tasks within the sensing regions of several hubs simultaneously, maximizing conflict but reducing travel time and stochastic execution uncertainty.}
     \label{fig:package-density}
 \end{figure}
Delivery requests are generated by sampling locations uniformly within a geographic bounding box, as shown in \cref{fig:package-density}. The window's start time $\underline{t}_k$ is drawn uniformly from $[t + w/2,\; t + w]$ and the window duration~$\overline{t}_k - \underline{t}_k$ from $[w/2,\, w]$, where $w$ is a nominal duration parameter. Unless otherwise specified, the simulation begins with $\lceil 1.5n \rceil$ initial packages, where $n$ is the number of agents. In dynamic-task experiments, new requests arrive at each time step independently with probability~$p_{\mathrm{new}}$.

\paragraph{\textbf{Travel time model}}
Following~\cite{dynamic-mr-ta}, travel time estimates between depot--delivery pairs are obtained from a precomputed matrix over a Halton point set covering the city, using nearest-neighbor lookup via a ball tree, and travel time uncertainty is modeled with the Epanechnikov distribution $\tau_{hk} \;\sim\; \mathrm{Epan}\!\bigl(\mu_{hk},\;\; \sigma_{hk})$ where $\mu_{hk}$ is the mean travel time from depot~$h$ to location~$\ell_k$, and $\sigma_{hk} = \mu_{hk}/3$ sets the standard deviation (scale factor~3.0). Outbound and return travel times are sampled independently. The success probability in~\eqref{eq:success-prob} is instantiated as $F_{\eta(i)k} = F_{\mathrm{Epan}}$, so that $\mathbb{P}[i \text{ completes } k \mid t] = F_{\mathrm{Epan}}(\overline{t}_k - t;\; \mu_{\eta(i)k},\; \sigma_{\eta(i)k})$.


 \begin{figure}
 \vspace{0.2cm}
     \centering
     \includegraphics[width=1\linewidth]{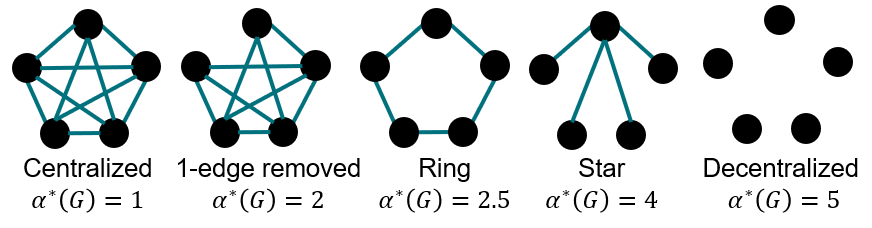}
     \caption{Communication Graph Topologies}
     \label{fig:graph-topologies}
 \end{figure}
\begin{figure*}
    \centering
    \begin{subfigure}[b]{0.245\textwidth}
        \includegraphics[width=\textwidth]{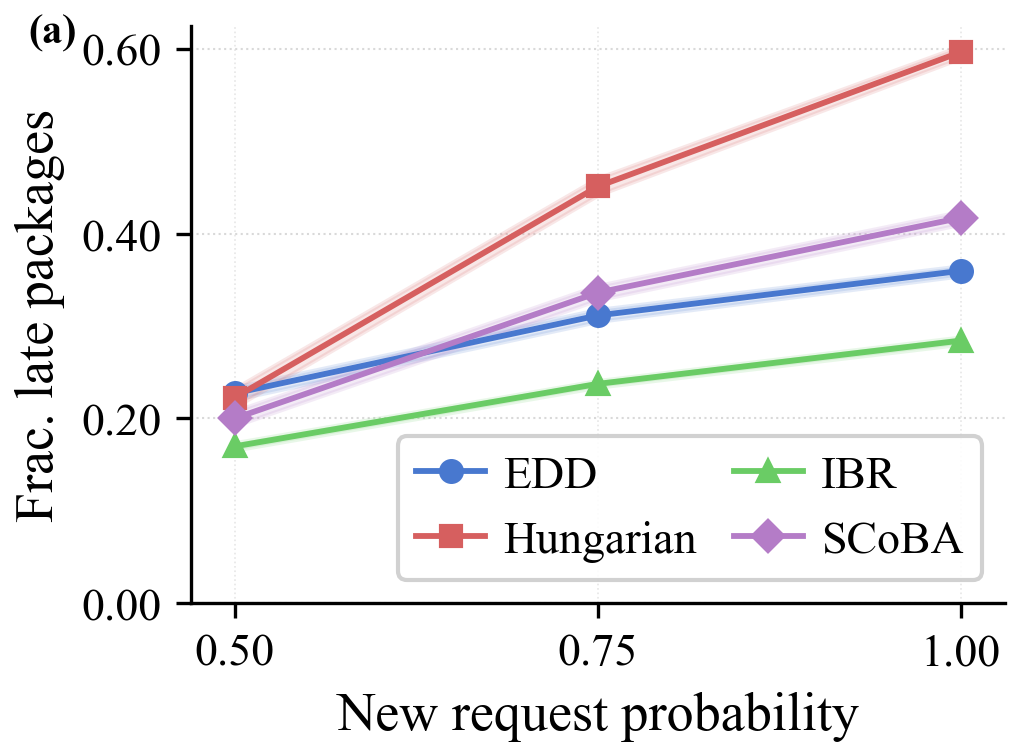}
        \label{fig:late-vs-requestprob}
    \end{subfigure}
    \hfill
    \begin{subfigure}[b]{0.245\textwidth}
        \includegraphics[width=\textwidth]{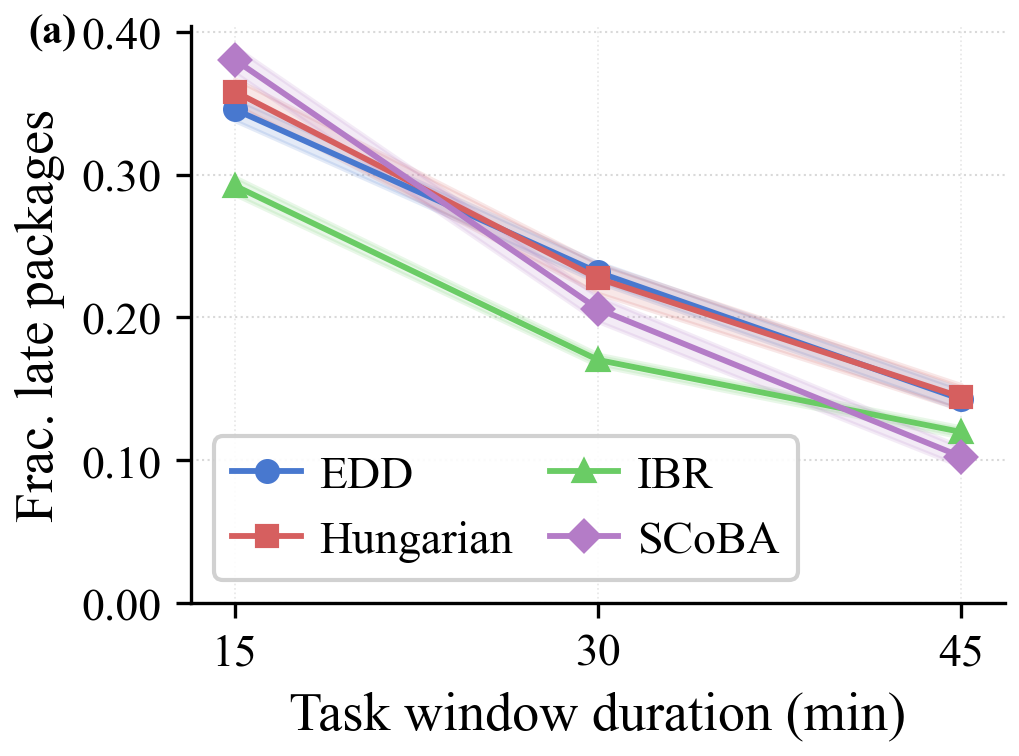}
        \label{fig:late-vs-taskwindow}
    \end{subfigure}
    \hfill
    \begin{subfigure}[b]{0.245\textwidth}
        \includegraphics[width=\textwidth]{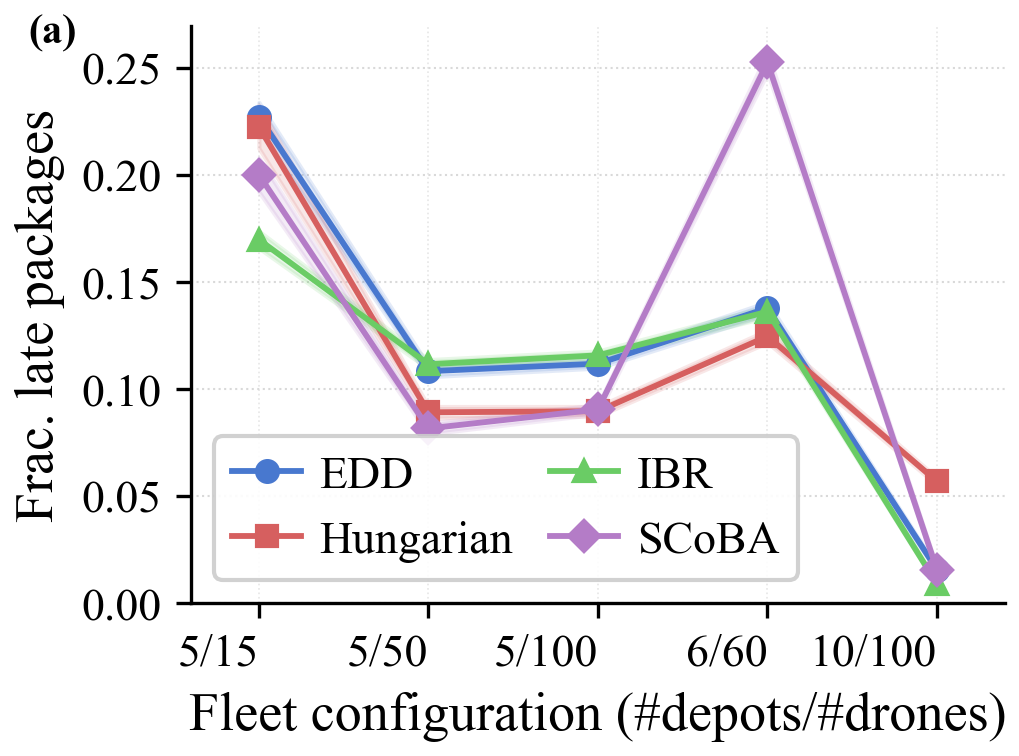}
        \label{fig:late-vs-dronenum}
    \end{subfigure}
    \hfill
    \begin{subfigure}[b]{0.245\textwidth}
        \includegraphics[width=\textwidth]{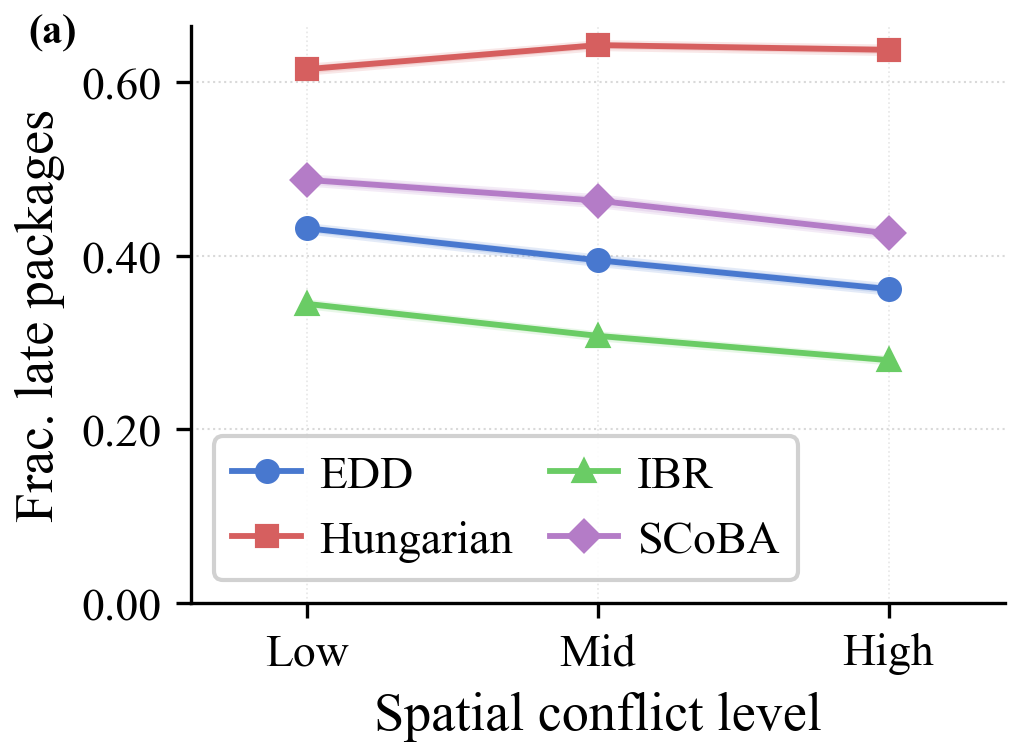}
        \label{fig:late-vs-density}
    \end{subfigure}

    \vspace{0.5em}

    \begin{subfigure}[b]{0.245\textwidth}
        \includegraphics[width=\textwidth]{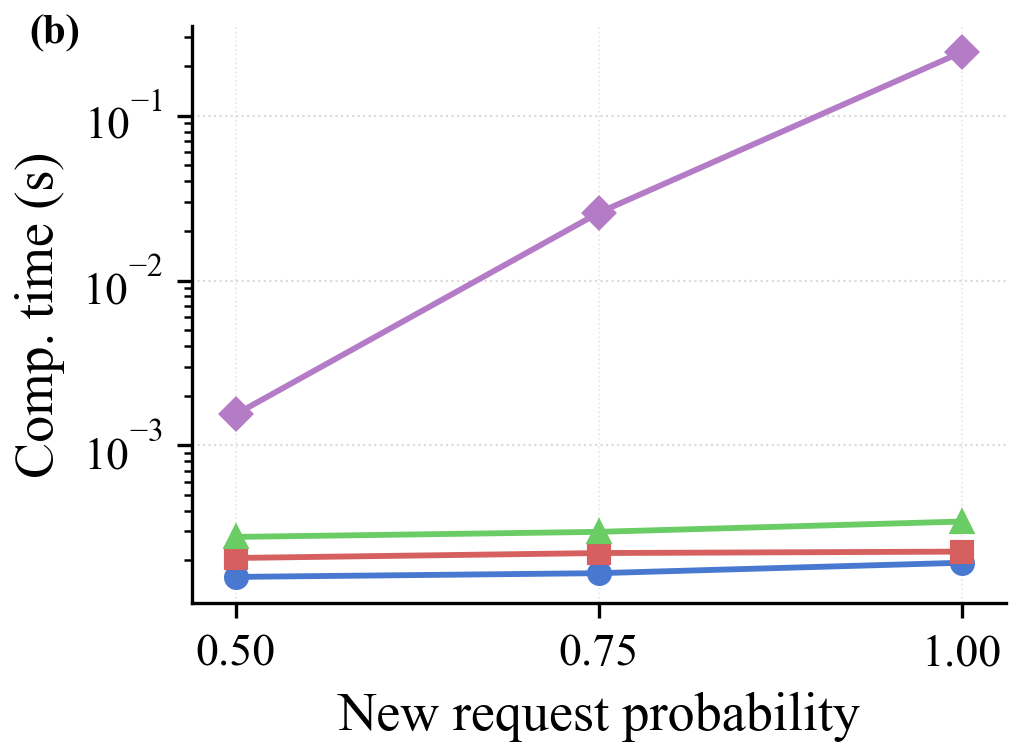}
        \caption{Request probability}
        \label{fig:time-vs-requestprob}
    \end{subfigure}
    \hfill
    \begin{subfigure}[b]{0.245\textwidth}
        \includegraphics[width=\textwidth]{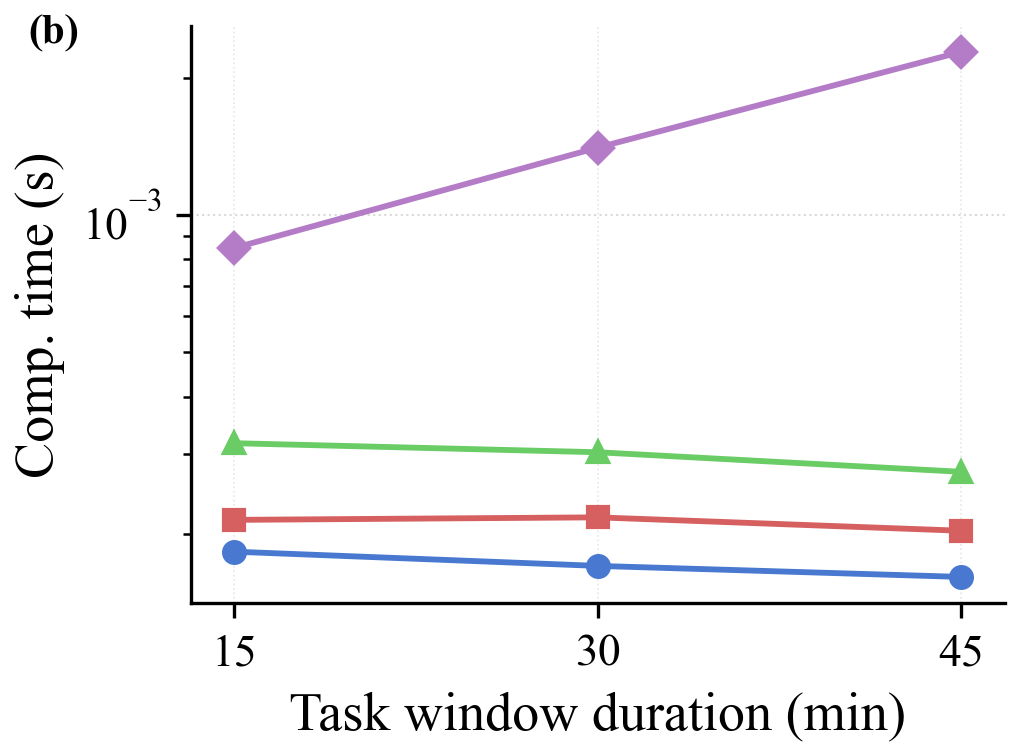}
        \caption{Service window}
        \label{fig:time-vs-taskwindow}
    \end{subfigure}
    \hfill
    \begin{subfigure}[b]{0.245\textwidth}
        \includegraphics[width=\textwidth]{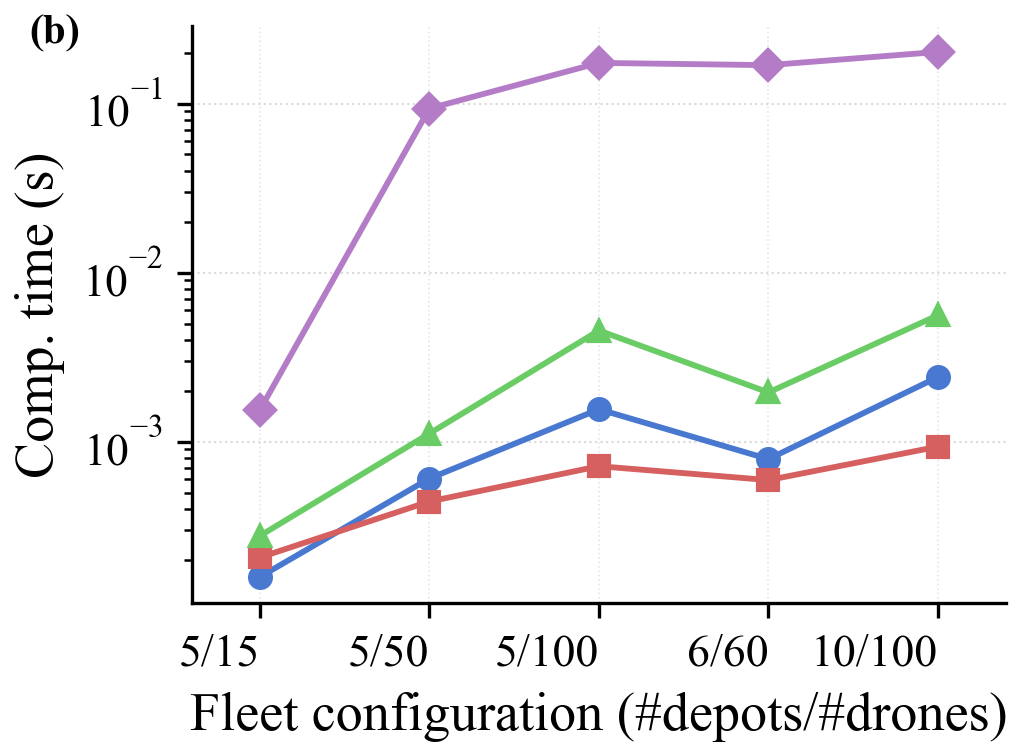}
        \caption{Fleet configuration}
        \label{fig:time-vs-dronenum}
    \end{subfigure}
    \hfill
    \begin{subfigure}[b]{0.245\textwidth}
        \includegraphics[width=\textwidth]{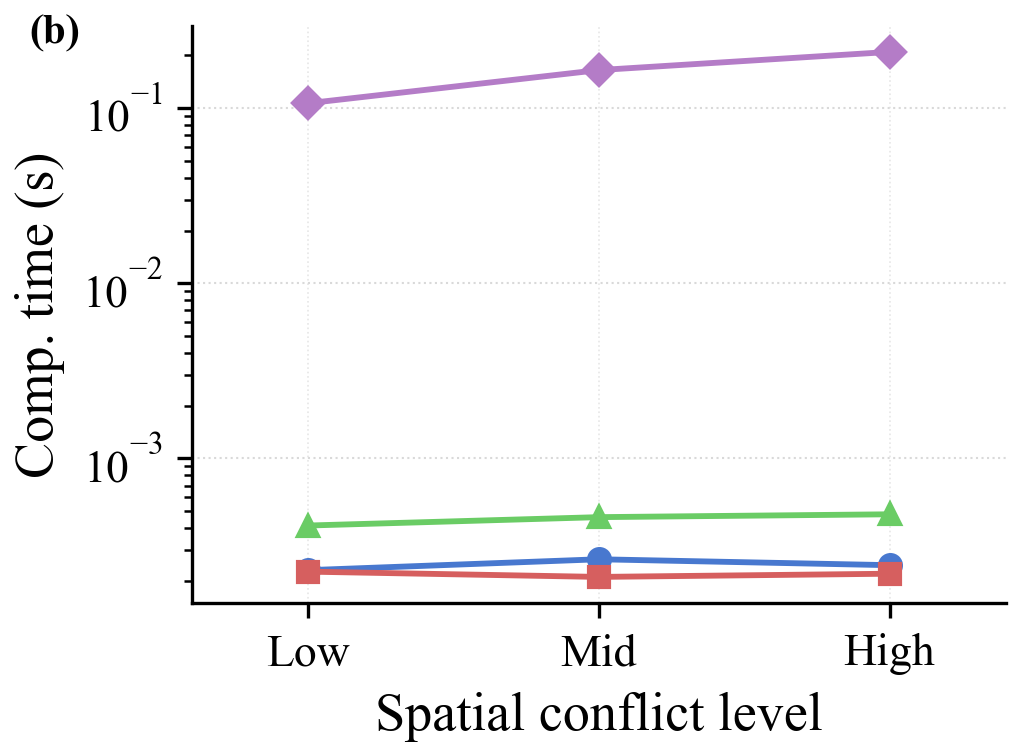}
        \caption{Spatial conflict level (Fig. \ref{fig:package-density})}
        \label{fig:time-vs-density}
    \end{subfigure}

    \caption{System performance across four algorithms under full communication, each column varying one parameter (100 trials each). \textbf{Top row:} mean fraction of late packages. \textbf{Bottom row:} avg.\ computation time per planning step (log scale).}
    \label{fig:full-comms-performance}
\end{figure*}

\subsection{Communication Graph Topologies} 
We evaluate performance under different communication graph topologies, shown in \cref{fig:graph-topologies}: a \textbf{centralized} (complete) graph, where every depot observes every other; a \textbf{star}, in which one depot connects bidirectionally to all others; a \textbf{ring}, a bidirectional cycle over depots in index order; a \textbf{decentralized} (empty) graph with no edges, so each depot operates in isolation; and \textbf{edge removal}.




\subsection{Baselines}
We compare IBR against the following baselines:
\begin{enumerate}
    \item \textbf{Earliest Due Date first (EDD)} assigns each agent to the task with the nearest time window deadline \cite[Chapter 3.2]{10.5555/1477600}.
    \item \textbf{Hungarian Algorithm} solves an agent–task assignment problem, where the weight of each agent–task pair corresponds to the success probability \cite{doi:10.1137/0105003}. 
    
    \item \textbf{Stochastic Conflict-Based Allocation (SCoBA)} is a tree-based method that constructs a policy by searching over possible assignments \cite{dynamic-mr-ta}.
\end{enumerate}
All three baselines are centralized by design, so to adapt them to our scenario, we impose an ordering over hubs (e.g., ascending or random) and let each hub allocate its agents in turn, committing at each step to tasks that no previous hub in its information group has already claimed. With a single group, this reduces to a single global assignment; as the number of groups grows, hubs commit based on partial views and duplicate assignments.

\subsection{Evaluation Metrics}\label{sec:eval-metrics}
We then evaluate the following metrics:
\paragraph{Fraction of late packages}
We define the fraction of late packages as
\begin{equation}
\text{Frac.\ Late}
  = 1 - \frac{1}{|\mathcal{K}|} \sum_{k \in \mathcal{K}} v_k(T),
\end{equation}
where $v_k(T) \in \{0,1\}$ indicates whether task~$k$ was completed by its deadline~$\overline{t}_k$, as defined in \cref{problem-form}.
A task is counted as late if it either expires without being attempted or is delivered after its deadline.
\paragraph{Computation time}
We report the average wall-clock time per planning step (i.e., per invocation of the allocation policy), displayed on a logarithmic scale.
\paragraph{Efficiency ratio}
We define the \emph{efficiency ratio} as the ratio of the empirical welfare attained by the IBR policy to the welfare achieved under full communication (centralized case). Thus, the ratio provides a normalized measure of how much performance degrades as the communication graph becomes sparser.

\section{Numerical Results}\label{sec:numerical-results}
Our experiments focused on two sets of simulations: a full communication baseline analysis (\cref{sec:full-comms}) and evaluation under different communication graph structures (\cref{sec:comms-results}). We implemented our simulations in Python and ran them on a commercial Lenovo ThinkPad laptop with Ubuntu~22.04.\footnote{Code is available at \href{https://github.com/sastry-group/Decentralization-Dynamic-Task-Assignment}{github.com/sastry-group/Decentralization-Dynamic-Task-Assignment.git}}

\subsection{Full-Communication Baseline}\label{sec:full-comms}

We first evaluate all four algorithms under full communication ($\alpha^*(G) = 1$) and then examine the effect of communication graph structure on IBR performance.
Each experiment uses 100 independent trials, and results are reported as the mean fraction of late packages. The default configuration uses 5~depots, 15~drones, new-request probability $p_{\mathrm{new}} = 0.5$, and service-window duration $w = 30$~minutes.
\paragraph{New request probability (\cref{fig:full-comms-performance}a)}
We vary $p_{\mathrm{new}} \in \{0.50, 0.75, 1.00\}$, controlling the rate at which new delivery requests appear per time step. As $p_{\mathrm{new}}$ increases, the task load grows relative to fleet capacity, and all algorithms exhibit higher late-delivery rates. IBR remains competitive across all settings, while EDD and SCoBA have comparable performance. 

\paragraph{Service-window duration (\cref{fig:full-comms-performance}b)}
We vary $w \in \{15,\, 30,\, 45\}$~minutes, determining the time each package remains available for delivery. Longer windows give drones more flexibility to reach distant packages, improving performance across all methods. IBR maintains the best performance across all window durations except for $w = 45$~min, where its performance approaches that of SCoBA.

\label{sec:comms-topologies}

\paragraph{Fleet configuration (\cref{fig:full-comms-performance}c)}
We evaluate depot/drone combinations $\{5/15,5/50, 5/100, 6/60,10/100\}$, varying both the number of depots and the drone-to-depot ratio. IBR and SCoBA achieve the best performance in most configurations; however, IBR's computation time is two orders of magnitude smaller than that of SCoBA.

\paragraph{Spatial conflict level (\cref{fig:full-comms-performance}d)}
We evaluate three spatial conflict levels---Low, Mid, and High---by adjusting the geographic bounding box from which delivery locations are sampled, as illustrated in \cref{fig:package-density}. In this experiment, the initial number of packages is set to 100 (compared to $1.5n$) and the service-window duration to 45~minutes. At the Low level, the sampling region is large relative to depot service areas, so packages are spread over a wide area; this reduces competition between depots for overlapping tasks, but increases mean travel times, since many packages lie near the edges of depot service ranges.
As the conflict level increases toward High, the sampling region shrinks and concentrates around the depots, reducing travel times but substantially increasing the fraction of packages that fall within the service range of multiple depots. This overlap creates more opportunities for redundant or conflicting assignments, particularly under limited communication. IBR continues to outperform the baselines across all conflict levels.
SCoBA also exhibits higher computation time at the High conflict level, consistent with the authors' observations in~\cite{dynamic-mr-ta} regarding the sensitivity of conflict-based search to the number of inter-agent conflicts.

Overall, under full communication, IBR matches the computation time of EDD and Hungarian while consistently completing more tasks. This is because IBR optimizes the welfare \eqref{eq: W(x, t)} directly, with each agent's utility defined as its marginal contribution. The baselines optimize proxies: EDD prioritizes urgency, Hungarian lets an earlier hub claim a task another hub would complete with higher probability, and SCoBA plans continuations that new arrivals can invalidate, which can be beneficial only under certain scenarios (e.g., longer service windows, \cref{fig:full-comms-performance}b, top). IBR's iterative updates instead reassign contested tasks to agents with higher success probability and allow redundant assignments when they increase expected completions.

\subsection{Effect of Communication Graph Structure}
\label{sec:comms-results}

To analyze performance under varying communication constraints, we compare EDD, Hungarian, IBR, and SCoBA across undirected graph topologies with increasing fractional independence number \eqref{eq: indep} from $\alpha^*(\mathcal{G})=1$ (centralized) to $\alpha^*(\mathcal{G})=5$ (decentralized), following the framework of \cite{grimsman-2019-impact}.

\begin{figure}
\vspace{0.2cm}
    \centering
    \includegraphics[width=\linewidth]{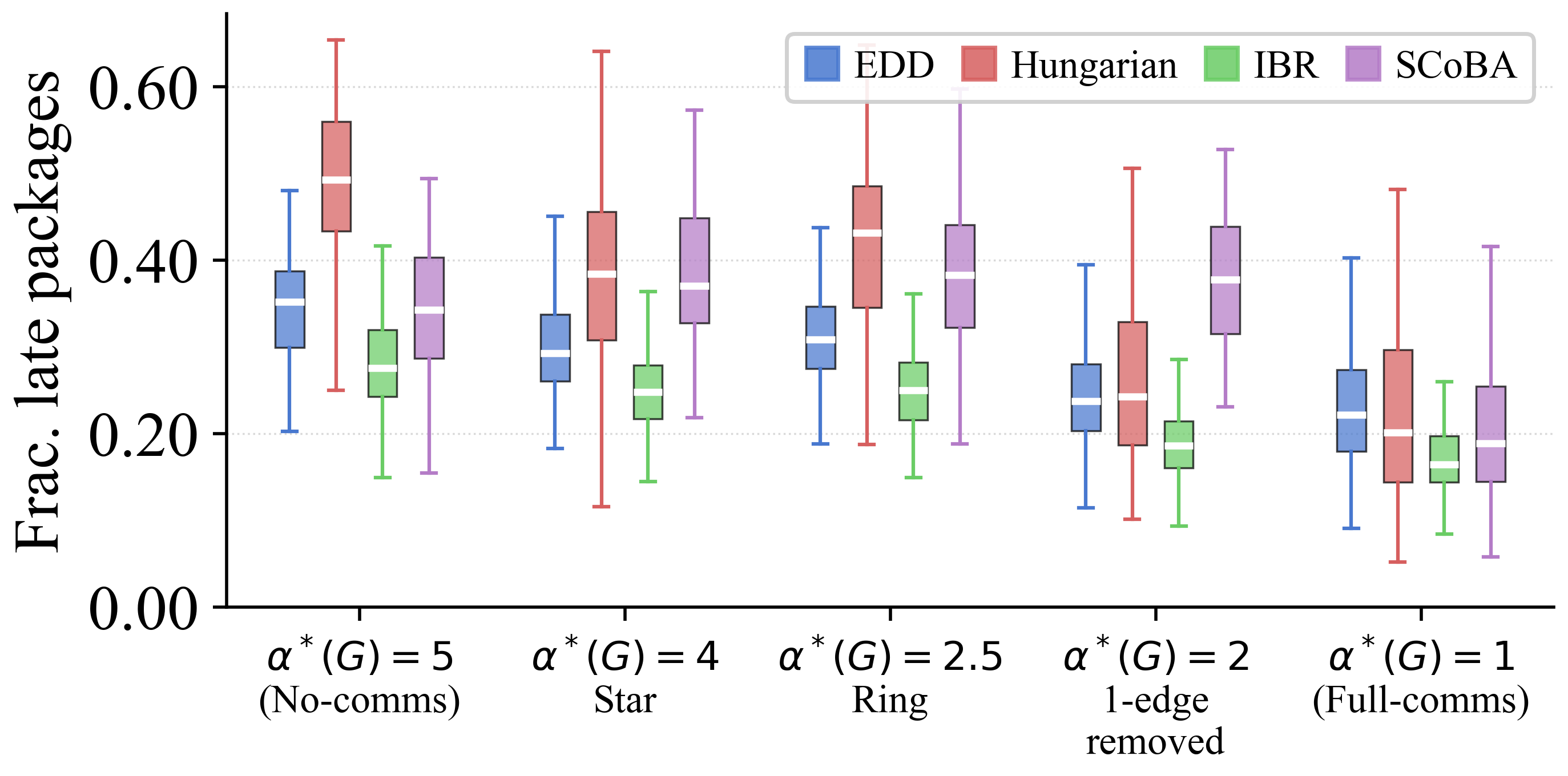}
    \caption{Distribution of the fraction of late packages across communication graph topologies (\cref{fig:graph-topologies}) for EDD, Hungarian, SCoBA, and IBR (100 trials, 5~depots, 15~drones)}
    \label{fig:box-plot-directed}
\end{figure}

\Cref{fig:box-plot-directed} shows that performance generally degrades as $\alpha^*(\mathcal{G})$ increases, but IBR maintains the lowest fraction of late packages and the smallest variance across all communication topologies. Because IBR does not require a fixed ordering of the agents, its allocation within each communication neighborhood is less sensitive to which hub acts first. SCoBA, in contrast, degrades under any non-full graph and behaves non-monotonically in $\alpha^*(\mathcal{G})$. This is because SCoBA relies on joint conflict-based search to deconflict agents' allocations; restricting this search to communication groups forces it to constrain agents within each group while conflicts across groups remain unresolved.

\begin{figure}
    \centering
    \includegraphics[width=1\linewidth]{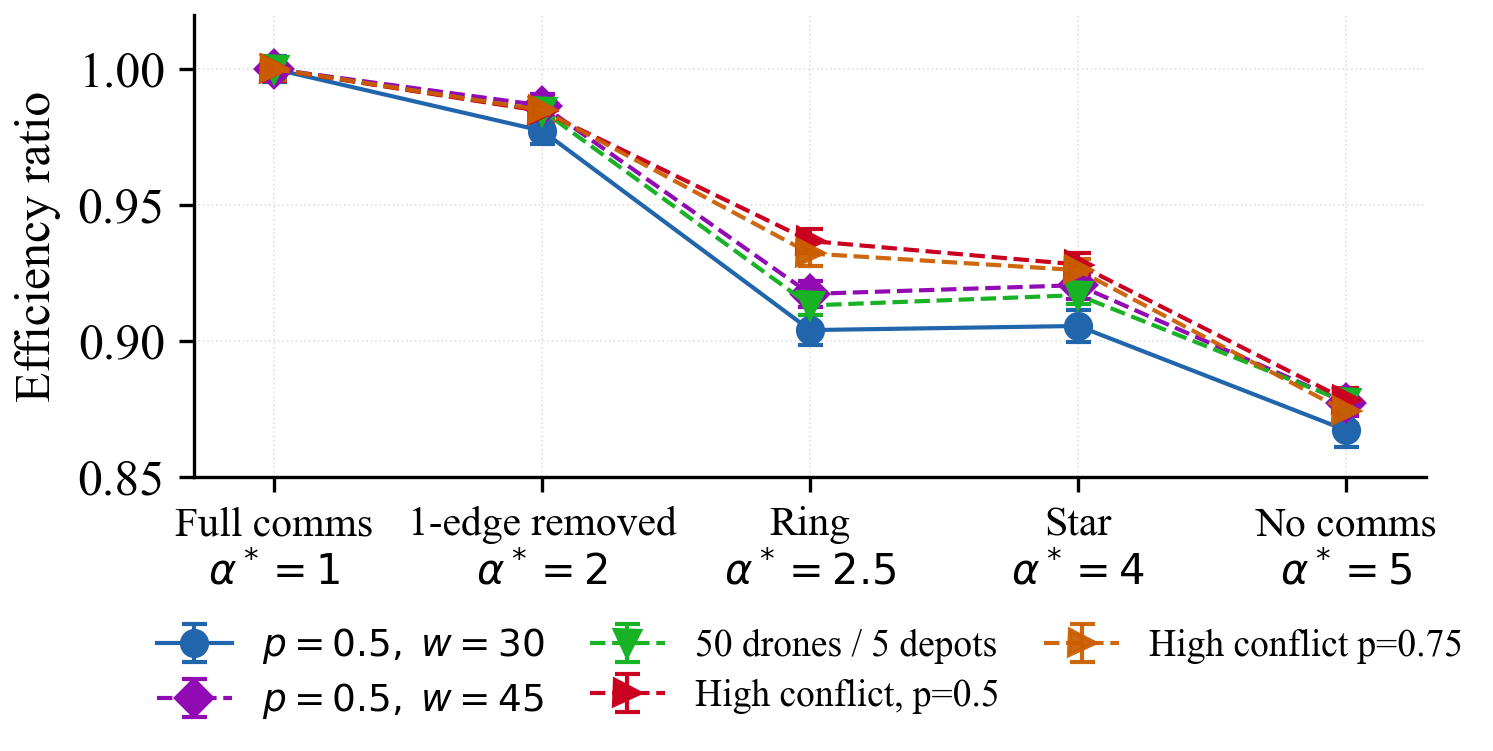}
    \caption{Efficiency ratio of IBR under different communication graph structures with different request probabilities $p$, service windows $w$, high spatial conflict with (5~depots, 15~drones) and (5~depots, 50~drones) configurations.}
    \label{fig:efficiency-ratio}
\end{figure}

\Cref{fig:efficiency-ratio} shows that removing a single edge ($\alpha^*(\mathcal{G})=2$) only reduces efficiency by $\sim2\%$ relative to full communication. Ring ($\alpha^*(\mathcal{G})=2.5$) and Star ($\alpha^*(\mathcal{G})=4$) graphs produce a sharp loss in efficiency to $\sim91-94\%$, and decentralization ($\alpha^*(\mathcal{G})=5$) yields   $\sim87\%$. These ratios align with the worst-case bound of \cref{prop: 1}.

\section{Conclusion}
In this work, we presented a framework for dynamic multi-robot task allocation under uncertain completion times, time-window constraints, and communication-limited decentralized execution. By modeling incomplete information through hub-based sensing regions and inter-hub communication graphs, we analyzed the trade-off between communication richness and coordination performance. We proposed Iterative Best Response (IBR) as a decentralized assignment policy and demonstrated empirically that it achieves task-completion rates competitive with centralized methods while incurring lower computational cost and retaining efficiency under moderate information loss.

Some limitations of our current model are that it assumes homogeneous agents, single-robot tasks, and fixed sensing ranges. In ongoing work, we are developing formal performance guarantees by comparing IBR outcomes against optimal allocations under dynamic task arrivals. This comparison will connect empirical performance to the price-of-anarchy bounds from the valid utility games framework \cite{Grimsman2-valid}. Further extensions include heterogeneous agent capabilities, time-varying communication topologies, tasks with heterogeneous rewards, and multi-robot task structures.


\bibliographystyle{IEEEtran} 
\bibliography{ref}

\addtolength{\textheight}{-12cm}   

\end{document}